\documentclass[a4paper,12pt]{article}
\usepackage[hdivide={*,15.5cm,*},vdivide={*,24cm,*} ,includefoot]{geometry}
\usepackage[latin1]{inputenc}
\usepackage[centertags]{amsmath}
\usepackage{amssymb,latexsym}
\usepackage[english,francais]{babel}
\usepackage{enumitem}
\usepackage{theorem}
\usepackage{color}
\normalbaselines
\def\ifm#1#2{\relax\ifmmode#1\else#2\fi}

\newcommand{\R}{\mathbb{R}}
\newcommand{\C}{\mathbb{C}}
\newcommand{\Z}{\mathbb{Z}}
\newcommand{\Q}{\mathbb{Q}}
\newcommand{\N}{\mathbb{N}}

\newcommand{\J}{\mathcal{J}}

\newcommand{\rk}{\text{rk}\,}
\newcommand{\sign}{\text{sign}\,}

\newcommand{\xon}    {\ifm {X_1,\ldots,X_n} {$X_1,\ldots,X_n$}}

\newcommand{\Fop}    {\ifm {F_1,\ldots,F_p} {$F_1,\ldots,F_p$}}
\newcommand{\iFop}    {\ifm {(F_1,\ldots,F_p)} {$(F_1,\ldots,F_p$)}}
\newcommand{\Fos}    {\ifm {F_1,\ldots,F_s} {$F_1,\ldots,F_s$}}

\newcommand{\Qxon}   {\ifm {\Q[X_1,\ldots,X_n]} {$\Q[X_1,\ldots,X_n]$}}

\newcommand{\eps}{\varepsilon}

\newcommand{\spar} {\vskip 0.05cm}

\newcommand{\be}{\begin{equation}}
\newcommand{\ee}{\end{equation}}

\newcommand{\bfs}{\boldsymbol}
\newcommand{\klk}{\ifm {,\ldots,} {$,\ldots,$}}

\newcommand{\kpk}{, \ldots ,}

\newenvironment{prf}{\vs\noindent \textbf {Proof.}}{$\mbox{}$\hfill $\Box$ \spar}

\def\cqfd{\vbox{\hrule height 5pt width 5pt }\bigskip}
\def\qed\cqfd
\def\noi{\noindent}
\def\vs{\smallskip}

\theoremstyle{break}

\newtheorem{proposition}{Proposition}

\newtheorem{theorem}[proposition]{Theorem}
\newtheorem{corollary}[proposition]{Corollary}
\newtheorem{lemma}[proposition]{Lemma}

\newcommand{\ol}{\overline}
\newcommand{\difrac}{\displaystyle\frac}
\usepackage{pdfsync}

\begin{document}
\selectlanguage{english}

\title{Intrinsic complexity estimates in polynomial optimization $^{1}$}

\author{Bernd Bank $^{2}\cdot$ Marc Giusti $^{3}\cdot$ Joos Heintz $^{4}\cdot$ Mohab Safey El Din $^{5}$}

\maketitle

\addtocounter{footnote}{1}\footnotetext{Research partially supported
  by the following Argentinian, French and Spanish grants: \\UBACYT 20020100100945, PICT--2010--0525,
Digiteo DIM 2009--36HD ``Magix'', ANR-2010-BLAN-0109-04 ``LEDA'',
GeoLMI, ANR 2011 BS03 011 06, EXACTA, ANR-09-BLAN-0371-01,
MTM2010-16051.
}
\addtocounter{footnote}{1}\footnotetext{Humboldt-Universit\"at zu
  Berlin, Institut f\"ur Mathematik, 10099 Berlin, Germany.\\
  bank@mathematik.hu-berlin.de}
\addtocounter{footnote}{1}\footnotetext{CNRS, Lab. LIX, \'Ecole
  Polytechnique, 91228 Palaiseau CEDEX, France.\\
  Marc.Giusti@Polytechnique.fr}
\addtocounter{footnote}{1}\footnotetext{Departamento de Computaci\'on,
  Universidad de Buenos Aires and CONICET, Ciudad Univ., Pab.I, 1428 Buenos Aires,
  Argentina, and Departamento de Matem\'aticas, Estad\'{\i}stica y
  Computaci\'on, Facultad de Ciencias, Universidad de Cantabria, 39071
  Santander, Spain.\newline joos@dc.uba.ar }
\addtocounter{footnote}{1}\footnotetext{Sorbonne Universities, Univ. Pierre et Marie Curie (Paris 06); INRIA Paris Rocquencourt, POLSYS Project; LIP6
CNRS, UMR 7606, Institut Universitaire de France.\\
 Mohab.Safey@lip6.fr}

\begin{center}

\bigskip

\end{center}

\begin{abstract}
\noindent It is known that point searching in basic semialgebraic sets and the search for globally minimal points in polynomial optimization tasks can be carried out using $(s\,d)^{O(n)}$ arithmetic operations, where $n$ and $s$ are the numbers of variables and constraints and $d$ is the maximal degree of the polynomials involved.\spar
\noindent Subject to certain conditions, we associate to each of these problems an intrinsic system degree which becomes in worst case of order $(n\,d)^{O(n)}$ and which measures the intrinsic complexity of the task under consideration.\spar
\noindent We design non-uniform deterministic or uniform probabilistic algorithms of intrinsic, quasi-polynomial complexity which solve these problems.
\end{abstract}

\bigskip


\section{Introduction}\label{s:0}

We develop uniform bounded error probabilistic and non--uniform
deterministic algorithms of intrinsic, quasi-polynomial complexity
for the point searching problem in basic semialgebraic sets and for
the search of isolated local and global minimal points in polynomial
optimization. The semialgebraic sets and optimization problems have
to satisfy certain well motivated geometric restrictions which allow
to associate with them an intrinsic {\it system degree} (see Section
\ref{ss:2.2}) that controls the complexity of our algorithms and
constitutes the core of their intrinsic character. The algorithms we
are going to design will become then polynomial in the length of the
extrinsic description of the problem under consideration and its
system degree (we take only arithmetic operations and comparisons in
$\Q$ into account at unit costs). The idea is that the system degree
constitutes a geometric invariant which measures the  intrinsic
``complexity'' of the {\it concrete} problem under consideration
(not of all problems like a worst case complexity). In worst case
the sequential  time complexity will be of order
$\binom{s}{p}(n\,d)^{O(n)}$ (respectively $(n\,d)^{O(n)}$), where $n$ is the
number of variables and $d$ the maximal degree of the polynomials
occurring in the problem description, $s$ their number and $1\le p
\le n$ the maximal codimension of the real varieties given by the
active constraints. We shall suppose that these polynomials are
represented as outputs of an essentially division--free arithmetic
circuit in $\Q[\xon]$ of size $L$ (here, we mean by essentially
division-free that only divisions by rational numbers are allowed).
The (sequential) complexity of our algorithms is then of order
$L\binom{s}{p}n^{O(p)}d^{O(1)}\delta^3$ (respectively
$L(n\,d)^{O(1)}\delta^3$), where $\delta$ is the intrinsic system
degree which in worst case becomes of order 
$(n\,d)^{O(n)}$. We call this type of complexity bounds {\it
intrinsic} and {\it quasi-polynomial}.
\medskip

For the problem of deciding the consistency of a given set of inequality constraints and of finding,
in case the answer is positive, a real algebraic sample point for each connected component  of the corresponding semialgebraic set, sequential time bounds of
simply exponential order, e.g. $(sd)^{O(n)}$, are exhibited in 
Grigor'ev and Vorobjov \cite{grivo}, Canny \cite{can}, Renegar \cite{rene1,rene2},
Heintz, Roy and Solern\'o \cite{hroy}, Basu, Pollack and Roy \cite{basu} and the 
book \cite{baporo}.
Such bounds can also be derived from efficient quantifier elimination algorithms over the reals (\cite{hroy,rene2,basu,baporo}). Since two alternating blocks of quantifiers become involved, one would expect at first glance that only a $(sd)^{O(n^2)}$  time complexity bound could be deduced from efficient real quantifier elimination for polynomial optimization problems. But, at least for global optimization, one can do much better with an $s^{2n+1}d^{O(n)}$ sequential time bound (see  \cite{baporo}, Algorithm 14.46). For particular global polynomial optimization problems the constant hidden in this bound can be made precise and
 the algorithms become implementable (see Greuet and Safey El Din \cite{gremo1,gremo},  Safey el Din \cite{Sa07,Sa08}, Greuet \cite{greu} and Jeronimo and Perucci \cite{jepe}). Accurate estimations for the minima are contained in \cite{jepet}. 
 The main difference with our approach is that these papers contain {\it extrinsic} worst case complexity bounds whereas our bounds are {\it intrinsic}.  
\medskip

Nevertheless, this article does not focus on the improvement of known worst case complexity bounds in optimization theory. Our aim is to exhibit classes of point searching problems in semialgebraic sets and polynomial optimization problems where it makes sense to speak about intrinsic complexity of solution algorithms. This is the reason why we put the accent on geometrical aspects of these problems. The algorithms become then borrowed from \cite{gilesa} (see also \cite{ghahmo,ghmo,hemawa,dule,cafma}) and, in particular, from \cite{bank3, bank4} (or alternatively from \cite{mosch1}). 

\subsection{Notions and Notations}\label{ss:01}
We shall freely use standard notions, results and notations from algebraic and semialgebraic geometry, commutative algebra and algebraic complexity theory which can be found e.g. in the books
\cite{mam,sha,mat,bcs}.\spar

Let $\Q$, $\R$ and $\C$ be the fields of the rational, real and complex
numbers, respectively, let $\xon$ be indeterminates
over $\C$ and let  $\Fop,\,1\le p \le n$, be polynomials of
$\R[\xon]$ defining
a closed, $\R$--definable subvariety $S$ of the
$n$--dimensional complex affine space $\C^n$.\spar

 We denote by $S_{\R}:=S\cap \R^n$ the real trace of the complex
variety $S$. We shall use also the following notations:
\[
\{F_1=0 \klk F_p=0\}:=S\;\;\text{and}\;\;\{F_1=0 \klk F_p=0\}_{\R}:=S_{\R}.
\]
For a given polynomial $Q\in \R[\xon]$ we denote by $S_Q$ and $(S_\R)_Q$ the  sets of points of $S$ and $S_\R$ at which $Q$ does not vanish
and call $S_Q$ the localization of $S$ at $\{Q=0\}$.\spar

We call a {\it regular} sequence $\Fop$  {\em reduced}
if for any index $1\le k\le p$ the ideal $(F_1\klk F_k)$ is radical.
A point $x$ of $\C^n$ is called {\em $\iFop$--regular} if
the Jacobian $J(F_1\kpk F_p) := \left[\frac{\partial F_j}{\partial X_k}\right]_ {{1 \le j \le p} \atop {1 \le k \le n}}$ has maximal rank $p$ at $x$. Observe, that for each {\em reduced} regular sequence $\Fop$ defining the variety $S$, the locus of $\iFop$--regular points of $S$ is the same. In this case we call an $\iFop$--regular point of $S$ simply {\em regular} (or {\em smooth}) or we say that $S$ is regular (or smooth) at $x$.
The variety $S$ is called $\iFop$--regular or smooth if $S$ is $\iFop$--regular at any of its points.\medskip 

Notice that the polynomials $\Fop$ form locally a reduced regular sequence at any $\iFop$--regular point of $S$.\medskip

Suppose for the moment that $V$ is a closed subvariety of $\C^n$. For $V$ irreducible we define its degree $\deg V$
as the maximal number of points we can obtain by cutting $V$ with finitely many affine hyperplanes of $\C^n$ such that the intersection is finite. Observe that this maximum is reached
when we intersect $V$ with dimension of $V$ many {\it generic} affine hyperplanes of $\C^n$.
In case that $V$ is not irreducible let $V=C_1\cup \cdots \cup C_s$ be the decomposition of $V$ into irreducible components.  We define the degree of $V$ as $\deg V:=\sum _{1\le j \le s}\deg C_j$.\spar

With this definition we can state the so-called {\it B\'ezout Inequality}:\spar
 Let $V$ and $W$ be closed subvarieties of $\C^n$. Then we have
\[\deg (V\cap W)\le \deg V\cdot \deg W.\]
If $V$ is a hypersurface of $\C^n$ then its degree equals the degree of its minimal equation. The degree of a point of $\C^n$ is just one. For more details we refer to  \cite{he,fu,vo}.
\medskip

Let $1\le i \le n-p$ and let $a:=\left[ a_{k,l}\right]_{{1 \le k \le
    n-p-i+1} \atop {0 \le l \le n}}$ be a real $((n-p-i+1)\times
(n+1)$--matrix with $(a_{1,0}\klk a_{n-p-i+1,0})\neq 0$
and suppose that $\left[ a_{k,l}\right]_{{1
    \le k \le n-p-i+1} \atop {1 \le l \le n}}$ has maximal rank
$n-p-i+1$.\spar

The $i$th dual polar variety of $S$ associated
with the matrix $a$ is defined as
closure of the locus of the $\iFop$--regular points of $S$ where all
$(n-i+1)$--minors of the polynomial $((n-i+1)\times
n)$--matrix
\[\scriptsize\begin{bmatrix}
\difrac{\partial F_1}{\partial X_1} & \cdots &
\difrac{\partial F_1}{\partial X_n}\\
\vdots & \vdots & \vdots \\
\difrac{\partial F_p}{\partial X_1} & \cdots &
\difrac{\partial F_p}{\partial X_n}\\
a_{1,1}-a_{1,0}X_1 & \cdots  & a_{1,n}-a_{1,0}X_n\\
\vdots & \vdots & \vdots \\
a_{n-p-i+1,1}-a_{n-p-i+1,0}X_1 & \cdots & a_{n-p-i+1,n}-a_{n-p-i+1,0}X_n\\
\end{bmatrix}
\]
vanish.\spar

Strictly speaking this notion of dual polar variety depends rather on the scheme given by the ideal generated by the polynomials $\Fop$ than on the variety $S$ itself. We shall not stick on the distinction between schemes and varieties, because it will be irrelevant in the sequel.
\medskip

Observe that this definition of
dual polar varieties may be extended to the case that there is given a
Zariski open subset $O$ of $\C^n$ 
and that $S$ is now the locally closed subvariety of $\C^n$ given by
\[
S:=\{F_1=0\klk F_p=0\}\cap O.
\]
In \cite{bank3} and \cite{bank4} we have introduced the
notion of dual polar variety of $S$ and motivated by
geometric arguments the calculatory definition above of these
objects. Moreover, we have shown that, for a real $((n-p-i+1)\times
(n+1))$--matrix $a=[a_{k,l}]_{1\le k \le n-p-i+1\atop{0\le l \le n}}$
with $[a_{k,l}]_{1\le k \le n-p-i+1\atop{1\le l \le n}}$
\emph{generic}, the $i$th dual polar variety is either empty or of pure codimension $i$ in
$S$. Further, we have shown that this polar variety
is normal and Cohen--Macaulay (but not necessarily smooth) at any of their $(F_1\klk F_p)$--regular points (see \cite{bghmum},
Corollary 2 and Section 3.1). This motivates the
consideration of the so--called {\em generic} dual polar varieties associated with real
$((n-p-i+1)\times (n+1))$--matrices $a$ which are generic in the above
sense, as invariants of the variety $S$.\spar

For our use of the word ``generic'' we refer to \cite{bghmum}, Definition 1.\spar

In case that $S$ is closed and that any point of  $S_{\R}$ is $\iFop$--regular, 
the $i$th dual polar variety associated with $a$
contains at least one point of each connected component of $S_{\R}$ and is therefore not empty
(see \cite{bank3} and \cite{bank4}, Proposition 2).\spar

If $S$ is only locally closed and $a$ is generic, then any $\iFop$--regular point of $S_\R$, which is a local minimizer of the distances of $(\frac{a_{1,1}}{a_{1,0}}\klk \frac{a_{1,n}}{a_{1,0}})$ to the points of $S_\R$, belongs to the $i$th dual polar variety of $S$ associated with $a$ (this fact is an immediate consequence of the proof \cite{bank3} and \cite{bank4}, Proposition 2).\spar

When speaking about generic dual polar varieties we shall always suppose that there is given a generic real or rational $(n-p)\times (n+1)$ matrix and that for $1\le i \le n-p$ the $i$th dual polar variety is associated with the first $n-p-i+1$ rows of this matrix. Hence our generic dual polar varieties will be arranged in descending chains.  
 
\subsection{Algorithmic tools}\label{ss:02}

In the sequel we shall make use of the Kronecker algorithm in the form  of \cite{gilesa}, Theorem 1 and 2 and of the following result which constitutes a reformulation and a slight sharpening of \cite{bank3}, Theorem 11 and \cite{bank4},
Theorem 13. This sharpening is obtained by applying \cite{bghlms}, Lemma 10 in the spirit of \cite{bghlms}, Section 5.1 to the original proofs.\spar

Let $Q, \Fop \in \Qxon, 1\le p \le n$, be polynomials with $Q\neq 0$ and $\deg F_j\le d,\; 1\le j \le p$. Assume that the polynomials $Q, \Fop$ are given as outputs of an essentially division--free circuit $\beta$ in $\Qxon$ of size $L$.

\begin{theorem}\label{t:0}
Let $\delta$ be the maximal degree of the Zariski closure of the $\iFop$--regular locus of $\{F_1=0\klk F_j=0\}_Q,\;1\le j \le p,$ and of all generic dual polar varieties of $S_Q=\{F_1=0\klk F_p=0\}_Q$.\spar

There exists a uniform bounded error probabilistic algorithm over $\Q$ which computes from the input $\beta$ in time $L(n\,d)^{O(1)}\delta^2\le (n\,d)^{O(n)}$
a representation by univariate polynomials of degree at most $\delta$ of a suitable,
over $\Q$--defined, $(n-p)$th generic dual polar variety of $S_\R$.\spar

For any $n,\,d,\,p,\,L,\,\delta \in \N$ with $1\le p \le
n$ this algorithm may be realized by an
algebraic computation tree over $\Q$ of depth
$L\,(n\,d)^{O(1)}\,\delta^2\le (n\,d)^{O(n)}$
that depends on certain parameters which are chosen randomly.
\end{theorem}

\spar

In view of the comments made at the end of  Subsection \ref{ss:01}, we may apply
the algorithm of Theorem \ref{t:0} in two ways to the problem of finding real algebraic sample points of $S_\R$.\medskip

The first way is to suppose $Q=1$ and that any real point of the closed variety 
$S=\{F_1=0\klk F_p=0\}$ is $\iFop$--regular. Then the algorithm returns a real algebraic sample for each connected component of $S_\R$. In Section \ref{s:1} we shall proceed in this manner.\spar

The second way works for locally closed varieties as well (i.e., in case $Q\notin \Q$)
and consists in the search for $\iFop$--regular real points of $S_Q$ which are local
minimizers of the distances of a suitable chosen point of $\R^n$ to the elements of the real trace of $S_Q$. In Subsection \ref{sss:2.2.2}  below we shall proceed in this manner in order to prove Theorem \ref{t:3}.

\section{Inequalities}\label{s:1}
Let $\Fos\in \R[\xon]$ and $1\le p \le \min\{s,n\}$. \medskip

{\bf Condition A}\spar

{\it Let $1\le j_1 < \cdots < j_k \le s,\;1\le k \le p$.
Then any point of the semialgebraic set $\{F_{j_1}=0\klk F_{j_k}=0\}_\R$ is  $(F_{j_1}\klk F_{j_k})$--regular. Moreover, any $p+1$ polynomials of $\Fos$ have no common real zero.}
\spar

Until the end of this section we shall tacitly assume that the polynomials $\Fos$ satisfy Condition A.\spar

For $\eps_1\klk \eps_s\in \{-1,1\}$ let
\[
\{\sign F_1=\eps_1\klk \sign F_s=\eps_s\}:=\{x\in \R^n\;|\;\sign F_1(x)=\eps_1\klk \sign F_s(x)=\eps_s\}.
\]
From now on we shall suppose without loss of generality $\eps_1=\cdots =\eps_s=1$ and write
\[
\{F_1>0\klk F_s>0\}\;\;\text{instead of}\;\;\{\sign F_1(x)=1\klk \sign F_s(x)=1\}.
\]
For the next two statements, let us fix a maximal index set $1\le j_1 < \cdots < j_k \le s,\;1\le k \le p$ with
\[
\{F_{j_1}=0\klk F_{j_k}=0\}_\R \cap \ol{\{F_1>0\klk F_s>0\}}\neq \emptyset.
\]
\begin{lemma}\label{l:0}~\vspace{-1cm}
\begin{multline*}
\{F_{j_1}=0\klk F_{j_k}=0\}_\R \cap \ol{\{F_1>0\klk F_s>0\}}=\\
=\{F_{j_1}=0\klk F_{j_k}=0\}_\R\cap \{F_j >0,\;1\le j \le s,\;j\neq j_1\klk j\neq j_k\}.
\end{multline*}
\end{lemma}
\begin{prf}
We show first the inclusion of the left hand side of the set equation in the right hand side. For this purpose, let $x$ be an arbitrary point of $\{F_{j_1}=0\klk F_{j_k}=0\}_\R \cap \ol{\{F_1>0\klk F_s>0\}}$. Suppose that there exists an index $1\le j \le s, \; j\neq j_1\klk j\neq j_k$ with $F_j(x)=0$. Then $x$ belongs to
$\{F_{j_1}=0\klk F_{j_k}=0, F_j=0\}_\R \cap \ol{\{F_1>0\klk F_s>0\}}$
which by the maximal choice of   $1\le j_1 < \cdots < j_k \le s$ is empty. Therefore, we have $F_j(x)\neq 0$ for any index $1\le j \le s, \; j\neq j_1\klk j\neq j_k$. Since $x$ belongs to $\ol{\{F_1>0\klk F_s>0\}}$, we have $F_j(x)>0$.\spar
We are now going to show the inverse inclusion. Consider an arbitrary point
$x\in \{F_{j_1}=0\klk F_{j_k}=0\}_\R\cap \{F_j >0,\;1\le j \le s,\;j\neq j_1\klk j\neq j_k\}$ and let $U$ be an arbitrary neighborhood of $x$ in $\R^n$. Without loss of generality we may assume that $U$ is contained in $\{F_j >0,\;1\le j \le s,\;j\neq j_1\klk j\neq j_k\}$. Since by Condition A the point $x$ is contained in the $(F_{j_1}\klk F_{j_k})$--regular set $\{F_{j_1}=0\klk F_{j_k}=0\}_\R$, the polynomial map from $\R^n$ to $\R^k$ given by $(F_{j_1}\klk F_{j_k})$ is a submersion at $x$ and therefore there exists a point $y\in U$ with
$F_{j_1}(y)>0\klk F_{j_k}(y)>0$. Because $U$ was an arbitrary neighborhood of $x$, we  conclude that $x$ belongs to $\{F_{j_1}=0\klk F_{j_k}=0\}_\R \cap \ol{\{F_1>0\klk F_s>0\}}$.
\end{prf}
\begin{corollary}\label{c:1}
Let $C$ be a connected component of $\{F_{j_1}=0\klk F_{j_k}=0\}_\R$ with
$C\cap \ol{\{F_1>0\klk F_s}$\\$\ol{>0\}}\neq \emptyset$. Then
\[C\subset \{F_j >0,\;1\le j \le s,\;j\neq j_1\klk j\neq j_k\}.\]
\end{corollary}
\begin{prf}
Lemma \ref{l:0} implies that the set
\[\{F_{j_1}=0\klk F_{j_k}=0\}_\R \cap \ol{\{F_1>0\klk F_s>0\}}\]
is open and closed in $\{F_{j_1}=0\klk F_{j_k}=0\}_\R$. Therefore this set is the union of all the connected components of $\{F_{j_1}=0\klk F_{j_k}=0\}_\R$ which have a nonempty intersection with $\ol{\{F_1>0\klk F_s>0\}}$. This implies Corollary \ref{c:1}.
\end{prf}
\subsection{Converting non-strict inequalities into strict ones}\label{ss:1.1}

Let $1\le k \le p,\;1\le j_1 < \cdots < j_k \le s $ and let $x=(x_1\klk x_n)$ be a point of
$\{F_{j_1}=0\klk F_{j_k}=0\}_\R \cap \{F_1>0\klk F_s>0\}$. Thus $x$ satisfies the system of non-strict inequalities
\[
F_{j_1}(x)\ge 0\klk F_{j_k}(x)\ge 0,\;\; F_j(x)>0,\;1\le j\le s,\;j\neq j_1\klk j\neq j_k
\]
Starting from $x$ we wish to construct a point $y\in \R^n$ which satisfies the strict inequalities
\[
F_j(y)> 0,\;1\le j \le s.
\]
From Condition A we conclude that the Jacobian $J(F_{j_1}\klk F_{j_k})$
has full rank $k$ at $x$. Therefore we may efficiently find a vector $\mu=(\mu_1\klk \mu_n)\in \R^n$ such that the entries of  $J(F_{j_1}\klk F_{j_k})(x)\mu^T$ are all positive (here, $\mu^T$ denotes the transposed vector of $\mu$).
\medskip

Let $Y$ be a new indeterminate and for $1\le j \le s$ let $G_j:=F_j(\mu_1 Y+x_1\klk \mu_n Y+ x_n)$. Observe, that the univariate polynomial $G_j$ satisfies the equation $\frac{dG_j}{dY}(0)=\sum_{1\le i \le n}\frac{\partial F_j}{\partial X_i}(x)\mu_i$. In particular, the entries of
\[
(\frac{dG_{j_1}}{dY}(0)\klk \frac{dG_{j_k}}{dY}(0))=J(F_{j_1}\klk F_{j_k})
(x)\mu^T
\]
are all positive. Let $c>0$ be the smallest positive zero of $\prod_{1\le j \le s}G_j$ (if there exists none, $c$ may be any positive real number). Then one verifies immediately that $z:=x+\frac{c}{2}\mu$ satisfies for any index $1\le j \le s$ the condition $F_j(z)=G_j(\frac{c}{2})>0$.

\subsection{Finding sample points for all consistent sign conditions}\label{ss:1.2}
Let $(\eps_1\klk \eps_s)\in\{-1,0,1\}^s$. The polynomial inequality system $\sign F_1=\eps_1\klk \sign $\\$F_s=\eps_s$ is called a {\it sign condition} on $\Fos$ which we say to be {\it consistent} if there exists a point $x\in \R^n$ satisfying it. In case
$(\eps_1\klk \eps_s)\in\{-1,1\}^s$ we call the sign condition {\it strict}, otherwise ${\it nonstrict}$. A real algebraic point of $\R^n$ which is supposed to be encoded ``\`a la Thom'' \cite{coro} and to satisfy the sign condition is called a {\it sample point} of it .\medskip

Let $\Fos$ be given as outputs of an essentially division--free arithmetic circuit $\beta$ in $\Q[\xon]$ (hence,
$\Fos$ belong to $\Q[\xon]$). Let $d\ge 2$ be an upper bound of $\deg F_1\klk \deg F_s$. For $1\le k \le p$ and $1\le j_1 < \cdots < j_k \le s$ let $\delta_{j_1\klk j_k}$
be the maximal degree of $\{F_{j_1}=0\klk F_{j_k}=0\}$ and all generic dual polar varieties of this variety. Let finally
\[
\delta:=\max\{\delta_{j_1\klk j_k}\;|\;1\le j_1 < \cdots < j_k \le s,\;1\le k \le p\}.
\]
We call $\delta$ the {\it degree} of the sample point finding problem for all consistent sign conditions of $\Fos$. From the B\'ezout Inequality we deduce
$\delta \le (nd)^{O(n)}$. Using Theorem \ref{t:0} we construct for each $1\le k \le p$ and $1\le j_1 <\cdots <j_k\le s$ real algebraic sample points for each connected component of $\{F_{j_1}=0\klk F_{j_k}=0\}_\R$. Then we evaluate the signs of all $F_j,\;1\le j \le s\;j\neq j_1\klk j\neq j_k$ on these sample points, which we think encoded ``\`a la Thom'' by univariate polynomials over $\Q$ of degree at most $\delta$. For this purpose we apply \cite{ros}, Proposition 4.9 (compare also \cite{culaped} and \cite{pedros}) at a computational cost of $O(\delta^3)$. 

By Corollary \ref{c:1} we obtain in this way  sample points for all non-strict consistent sign conditions on $F_1\klk F_s$. As far as only sample points for the strict sign conditions on $F_1\klk F_s$ are required, we limit our attention to sample points of the connected component of $\{F_{j_1}=0\klk F_{j_k}=0\}_\R$ where  the signs of all $F_j, 1\le j \le s\;\;j\neq j_1 \klk j \neq j_k$ are all strict. Let $x$ be such a sample point with $\sign F_j(x)=\eps_j$, and $\eps_j\in\{-1,1\}$ for $1\le j \le s$ with $j\neq j_1\klk j\neq j_k $.\spar

Then, following Subsection \ref{ss:1.1},  for any $\eps_{j_1}\klk
\eps_{j_k}\in \{-1,1\}$ we may convert $x$ into a real algebraic
sample point of the strict sign conditions $\sign F_1=\eps_1\klk
\sign F_s=\eps_s$. The whole procedure can be realized in time
$L\binom{s}{p}\,n^{O(p)}\,d^{O(1)}\,\delta^3$ (here arithmetic
operations and comparisons in $\Q$ are taken into account at unit
costs). We have therefore shown the following statement which
constitutes a simplified variant of  \cite{legum}, Theorem 5.

\begin{theorem}\label{t:1}
Let $n,\,d,\,p,\,s,\,L,\,\delta\in \N$ with $1\le p \le n$ be
arbitrary and let $F_1\klk F_s\in \Q[\xon]$ be polynomials of degree
at most $d$ satisfying Condition A and having sample point finding
degree at most $\delta$. Suppose that $F_1\klk F_s$ are given as
outputs of an essentially division-free circuit $\beta$ in
$\Q[\xon]$ of size $L$.\medskip

There exists a uniform bounded error probabilistic algorithm
$\mathcal{A}$ over $\Q$ which computes from the input $\beta$ in
time
$L\binom{s}{p}\,n^{O(p)}\,d^{O(1)}\,\delta^3\le \binom{s}{p}\,(nd)^{O(n)}$
real algebraic sample points for each consistent sign condition on
$F_1\klk F_s$.
\medskip

For any $n,\,d,\,p,\,s,\,L,\,\delta \in \N$ with $1\le p \le
n$ the probabilistic algorithm $\mathcal{A}$ may be realized by an
algebraic computation tree over $\Q$ of depth
$L\,\binom{s}{p}\,n^{O(p)}\,d^{O(1)}\,\delta^3\le \binom{s}{p}(n\,d)^{O(n)}$
that depends on certain parameters which are chosen randomly.
\end{theorem}

Condition A requires that for every  $1\le k \le p$ and $1\le j_1 <\cdots <j_k\le s$
any point of the semialgebraic set $\{F_{j_1}=0\klk F_{j_k}=0\}_\R$ is $(F_{j_1}\klk F_{j_k})$--regular. This requirement may be relaxed using the algorithmic tools developed in \cite{bghlms}. More restrictive is the requirement that any $p+1$ polynomials of $\Fos$ have no common real zero. If we drop this requirement, we have to modify the notion of the degree $\delta$ of the sample point finding problem for all consistent sign conditions of $\Fos$. We obtain then a complexity bound of order $L\binom{s}{n}\,n^{O(n)}\,d^{O(1)}\,\delta^3\le \binom{s}{n}\,(nd)^{O(n)}$. The exponential behavior of the ``combinatorial'' complexity $\binom{s}{n}\,n^{O(n)}$ cannot be avoided, since for $\Fos$ being generic polynomials of degree one at least $\binom{s}{n}\,2^n$ distinct sign conditions become satisfied.

\section{Optimization}\label{s:2}
We associate with a polynomial optimization problem with smooth equality constraints certain natural geometric conditions and an intrinsic invariant that controls the complexity of the algorithm which we are going to develop in order to solve this problem. Our approach has some features in common with that of \cite{ggmz}.

\subsection{Geometric considerations}\label{ss:2.1}

Let be given polynomials $G, \Fop\in \R[\xon],\;1\le p \le n$, and let 
$V:=\{F_1=0\klk F_p=0\}$. We
suppose for the rest of this paper that $V_\R$ 
is not empty and that any point of $V_\R$ is $\iFop$--regular.\spar

From now on, let $\J$ denote an ordered sequence of indices $j_1\klk j_p$ with $1\le j_1< \cdots < j_p\le n$.\spar

For such an {\it index sequence} $\J$ denote by $\Delta_\J$ the $p$--minor of the Jacobian $J(\Fop)$ given by the columns numbered by the elements of $\J$ and by
\[
M_1^\J\klk M_{n-p}^\J
\]
the $(p+1)$--minors of $((p+1)\times n)$--matrix
\[
\begin{bmatrix}
J(\Fop)\\ \frac{\partial G}{\partial X_1}\cdots  \frac{\partial G}{\partial X_n}
\end{bmatrix}
\]
given by the columns numbered by the elements of $\J$ to which we add, one by one, the columns numbered by the indices belonging to the set $\{1\klk n\}\setminus \J$.\spar

Let $x$ be a local minimal point of $G$ on $V_\R$. Then the Karush-Kuhn-Tucker conditions
imply that
$\rk \begin{bmatrix}
J(\Fop)\\ \frac{\partial G}{\partial X_1}\cdots  \frac{\partial G}{\partial X_n}
\end{bmatrix}(x)\le p$
holds. We consider the subset $W$ of $V$ satisfying this rank condition, i.e.,
\[
W:=\{x\in V\;|\; \rk \begin{bmatrix}
J(\Fop)\\ \frac{\partial G}{\partial X_1}\cdots  \frac{\partial G}{\partial X_n}
\end{bmatrix}(x)\le p\}.
\]
Let $\J$ be fixed and let us consider the localization $W_{\Delta_{\J}}$ of $W$ outside of the hypersurface $\{\Delta_{\J}=0\}$.
In \cite{bank2} we developed a succinct local description of determinantal varieties. The  main tool for this description was a
general Exchange Lemma which depicts an exchange relation between certain minors
of a given matrix . Applying this Exchange Lemma we conclude
\[
W_{\Delta_\J}=V_{\Delta_\J}\cap \{M_1^\J=0\klk M_{n-p}^\J=0\}.
\]
For the rest of this paper we shall assume that the polynomials $G$ and $\Fop$ satisfy the following condition.\medskip

Let $D_k$ denote the union of all irreducible components of $W$ of dimension strictly larger than $n-p-k$. Observe that $D_{n-p+1}$ is also well defined.
\medskip

{\bf Condition B}\spar

{\it Let $\J$ be an arbitrary index sequence, $1\le k \le n-p$ and  denote by $C_1^\J\klk C_{s_\J}^\J$ the irreducible components of $(V_{\Delta_\J}\cap \{M_1^\J=0\klk M_k^\J=0\}) \setminus D_k$. Then any point of
\[
\left((V_\R)_{\Delta_\J}\cap \{M_1^\J=0\klk M_k^\J=0\}_\R\right)\setminus (D_k)_\R
\]
is $(\Fop, M_1^\J\klk M_k^\J)$--regular. Moreover for any $1\le j \le s_\J$ the semialgebraic set $(C_j^\J)_\R$ is non-empty and each irreducible component of $W$ is of dimension strictly smaller than  $n-p$ and contains a real point.}
\medskip

In particular, Condition B implies that the real trace of $V_{\Delta_\J}\cap \{M_1^\J=0\klk M_k^\J=0\}\setminus D_k$ is smooth at any of its points .
This entails that for $1\le i< j \le s_\J$ the semialgebraic sets $(C_i^\J)_\R$ and $(C_j^\J)_\R$ have an empty intersection.\medskip

Condition B allows to establish a bridge between semialgebraic and algebraic geometry (see below the proofs of Lemma \ref{l:1} and \ref{l:2}).
\medskip

\begin{lemma}\label{l:1}
Let notations be as in Condition B, which we suppose to be satisfied, and let $\J$ be an arbitrary index sequence.  Then for 
any $1\le j \le s_\J$, we have $\dim C_j^\J=n-p-k$.
\end{lemma}
\begin{prf}
From Condition B we infer that there exists  an open semialgebraic subset $U$ of $\R^n$, disjoint from $(D_k \cup C_1^\J\cup\, \cdots\, \cup C_{j-1}^\J\cup C_{j+1}^\J\cup \, \cdots \, \cup C_{s_\J}^\J\})_\R$, with $U\cap (C_j^\J)_\R\neq \emptyset$. This implies
\[
U\cap (V_\R)_{\Delta_\J}\cap \{M_1^\J=0\klk M_k^\J=0\}_\R=U\cap (C_j^\J)_\R.
\]
From $U\cap (D_k)_\R=\emptyset$ and Condition B we deduce now that any point
of the non--empty semialgebraic set $U\cap (C_j^\J)_\R$ is $(\Fop, M_1^\J\klk M_k^\J)$--regular. In particular, $C_j^\J$ is included in $\{F_1=0\klk F_p=0, M_1^\J=0\klk M_k^\J=0\}$ and contains a $(\Fop, M_1^\J\klk M_k^\J)$--regular point. This implies $\dim C_j^\J\le n-p-k$. From the definition of $C_j^\J$ we infer $\dim C_j^\J\ge n-p-k$. Therefore we have $\dim C_j^\J = n-p-k$.
\end{prf}
\begin{lemma}\label{l:2}
Suppose that Condition B is satisfied and let $C$ be an irreducible component of $W$. Then $G$ takes a constant real value on $C$.
\end{lemma}
\begin{prf}
Since by assumption $V_\R=\{F_1=0\klk F_p=0\}_\R$ is smooth and since $C_\R$ is nonempty by Condition B,  there exists an index sequence $\J$  with $(C_\R)_{\Delta_\J}\neq \emptyset$. Furthermore, there exists an index $1\le k\le n-p$ with $\dim C =n-p-k$.
\spar

By Lemma \ref{l:1}, we may assume without loss of generality that
$C_{\Delta_\J}\setminus D_k=C_1^\J$ holds. Let $x$ be an arbitrary point of $(C_1^\J)_\R$. As we have seen in the proof of Lemma \ref{l:1}, there exists an open semialgebraic neighborhood $U$ of
$x$ in $\R^n$ with
\[
U\cap (V_\R)_{\Delta_\J}\cap \{M_1^\J=0\klk M_k^\J=0\}_\R=U\cap (C_1^\J)_\R
\]
and $U\cap (D_k)_\R= \emptyset$. Condition B implies now that $U\cap (C_1^\J)_\R$ is a smooth semialgebraic manifold which we may suppose to be
connected by continuously differentiable paths.  Let $y$ be an arbitrary point of $U\cap (C_1^\J)_\R$ and let $\tau$ be a continuously differentiable path in $U\cap (C_1^\J)_\R$ that connects $x$ with $y$. We may suppose that $\tau$ can be extended to a suitable open neighborhood of $[0,1]$ in $\R$ and that $\tau(0)=x$ and $\tau(1)=y$ holds. Observe that $\tau([0,1])$ is contained in $V_\R$.
The path $\tau$ depends on a parameter $T$ defined in the given neighborhood of $[0,1]$. Let $\tau(T):=(\tau_1(T)\klk \tau_n(T))$. Since $\tau([0,1])$ is contained in $V_\R$ the vector $(\frac{d\tau_1}{dT}(t)\klk \frac{d\tau_n}{dT}(t))$ belongs to the kernel of $J(\iFop)(\tau(t))$ for any $t\in [0,1]$. On the other hand, $C_1^\J\subset C \subset W$ implies that
$(\frac{\partial G}{\partial X_1}(\tau(t))\klk \frac{\partial G}{\partial X_n}(\tau(t))$ is linearly dependent on the full rank matrix $J\iFop(\tau(t))$.
Therefore we have
$\frac{d(G\circ \tau)}{dT}(t)=\sum_{i=1}^n \frac{\partial G}{\partial X_i}(\tau(t))\frac{d\tau_i}{dT}(t)=0$ for any $t\in [0,1]$. Hence, $G\circ \tau$
is constant on $[0,1]$. Consequently, we have $G(x)=G(\tau(0))=G(\tau(1))=G(y)$. From the arbitrary choice of $x$ and $y$ in $U\cap (C_1^\J)_\R$ we infer that $G$ takes on $U\cap (C_1^\J)_\R$ a constant value. Thus the restriction of $G$ to the semialgebraic set
$(C_1^\J)_\R$ is locally constant and takes therefore only finitely many values in $\R$. \spar

By Condition B there exists an $(\Fop, M_1^\J\klk M_k^\J)$--regular point $x=(x_1\klk x_n)$ of $(C_1^\J)_\R$. Hence, there exists an open semialgebraic neighborhood $U'$ of $x$ in $\R^n$ and $n-p-k$ parameters $\xi_1\klk \xi_{n-p-k}$ of $U'\cap C_1^\J$ such that the restriction of $G$ to $U'\cap (C_1^\J)_\R$ can be developed into a convergent power series $P(\xi_1-x_1\klk \xi_{n-p-k}-x_{n-p-k})$ around $(x_1\klk x_{n-p-k})$. Since $G$ is locally constant on $U'\cap (C_1^\J)_\R$ we conclude that $P(\xi_1-x_1\klk \xi_{n-p-k}-x_{n-p-k})$ equals its constant term, say $b\in \R$. \spar

On the other hand, there exists an open neighborhood $O$ of $x$ in $\C^n$ such that the restriction of $G$ to $O\cap C_1^\J$
can be developed into a convergent power series in $\xi_1-x_1\klk \xi_{n-p-k}-x_{n-p-k}$. This power series must necessarily be $P(\xi_1-x_1\klk \xi_{n-p-k}-x_{n-p-k})$. Thus
$G$ takes on $O\cap C_1^\J$ only the real value $b$. Suppose that $G$ takes on $C_1^\J$ a value different from $b$. Then $(C_1^\J)_{G-b}$ is nonempty and therefore (by \cite{mam}, Ch. I, Section 10, Corollary 1) dense in the Euclidean topology of $C_1^\J$.  In particular,
there exists a point $y\in O\cap C_1^\J$ with $G(y)\neq b$. This contradiction implies that $G$ takes on $C_1^\J$ the constant value $b$. Lemma \ref{l:2}
follows now from the fact that $C_1^\J$ is dense in $C$.
\end{prf}
\medskip

A formally different result of the same spirit as Lemma \ref{l:2} is \cite{denipo}, Lemma 3.3. Its proof can be transformed into an alternative argument for Lemma \ref{l:2}.
\medskip

Let $1\le k \le n-p$. By Lemma \ref{l:2} the polynomial $G$ takes on $D_k$ only finitely
many values which are all real algebraic. We denote $B_k:=G(D_k)$ the set of these values.
\medskip

To any index sequence $\J$ we may associate a Hessian matrix $H_\J$ of $G$ on $V_{\Delta_\J}$ whose entries belong to $\R[\xon]_{\Delta_\J}$. The following condition reflects the intuitive meaning of the Hessian.\spar

{\bf Condition C}\spar

{\it Let $\J$ be an arbitrary index sequence. Then the rational function $\det H_\J$ does not vanish at any $(F_1\klk F_p, M_1^\J\klk M_{n-p}^\J)$--regular real point of $W_{\Delta_\J}$.}

\begin{lemma}\label{l:4}
Suppose that Conditions B and C are satisfied and let $\J$ be an arbitrary index sequence. Then the set of isolated local minimal points of $G$ on $(V_\R)_{\Delta_\J}$ is exactly the set of $(F_1\klk F_p, M_1^\J\klk M_{n-p}^\J)$--regular points of $(W_\R)_{\Delta_\J}$ where $H_\J$ is positive definite.
\end{lemma}
\begin{prf}
Let $\J$ be an arbitrary index sequence.
From the Morse Lemma (see \cite{dem}) one deduces easily that the points of $(W_\R)_{\Delta_\J}$ where $H_\J$ is positive definite, are isolated local minimizers of $G$ on $(V_\R)_{\Delta_\J}$. So, we have only to show that the isolated local minimal points of $G$ in $(V_\R)_{\Delta_\J}$ belong to $(W_\R)_{\Delta_\J}$, 
are $(F_1\klk F_p, M_1^\J\klk M_{n-p}^\J)$--regular and that their Hessians are positive definite.\spar

Let $x\in (V_\R)_{\Delta_\J}$ be an isolated minimal point of $G$ in $(V_\R)_{\Delta_\J}$. Then, as we have seen, $x$ belongs to $(W_\R)_{\Delta_\J}$. Let $C$ be an arbitrary irreducible component of $W_{\Delta_\J}$ which contains $x$. Let $n-p-k$ with $1\le k \le n-p$ be the dimension of $C$. Suppose that $1\le k <n-p$ holds. Condition B implies now that there exists an open subset of
 $(\Fop, M_1^\J,$ $\ldots , M_k^\J)$--regular points of $C_\R$ which is dense in $C_\R$. This implies that any neighborhood of $x$ in $C_\R$
contains a point $y$ different from $x$. Since $G(y)=G(x)$ holds by Lemma \ref{l:2}, the local minimal point $x$ of $G$ in $(V_\R)_{\Delta_\J}$ cannot be isolated. Therefore, we have $k=n-p$.  From Condition B we deduce now that $x$ is $(\Fop, M_1^\J,$ $\ldots , M_{n-p}^\J)$--regular. Hence, by Condition C, we have $\det H_\J(x)\neq 0$. The Morse Lemma implies now that  $H_\J(x)$
must be positive definite for $x$ being an isolated local minimal point of $G$ in
$(V_\R)_{\Delta_\J}$.
 \end{prf}
\spar
Finally let us comment the regularity requirement contained in Condition B by two classes of examples.\spar

Let $(a_1\klk a_n)\in \R^n$ be a generic vector and let $G:=a_1X_1+\cdots +a_nX_n$ or $G:=(a_1-X_1)^2+\cdots +(a_n-X_n)^2$. Furthermore, let $\Fop$ be as at the beginning of this subsection. Mimicking the argumentation of \cite{bghmum}, Section 3 we see that for any index sequence $\J$ and any index $1\le k \le n-p$
every point of the real trace of $V_{\Delta_\J}\cap \{M_1^\J=0\klk M_k^\J=0\}$
is $(\Fop, M_1^\J\klk M_k^\J)$--regular. Hence the regularity requirement contained in Condition B becomes satisfied for this kind of examples.
\subsubsection {Unconstrained optimization}\label{sss:1}

We illustrate our argumentation in the case of unconstrained optimization. In this case we have $p:=0$ and $V$ is the complex affine space $\C^n$. There is given a polynomial
$G\in \R[\xon]$ and the task is to characterize the isolated local and global minimal points of $G$ in $\R^n$.
Such a local minimal point belongs to $W:=\{\frac{\partial G}{\partial X_1}=0\klk \frac{\partial G}{\partial X_n}=0\}$. For the unconstrained optimization problem we consider the following condition.\spar

Let $D_k$ be the union of all irreducible components of $W$ of dimension strictly larger  than $n-k$.\medskip

{\bf Condition D}\spar

{\it Let $1\le k \le n$ and let $C_1\klk C_s$ be the irreducible components of $\{\frac{\partial G}{\partial X_1}=0\klk \frac{\partial G}{\partial X_k}=0\}\setminus D_k$. Any point of $\{\frac{\partial G}{\partial X_1}=0\klk \frac{\partial G}{\partial X_k}=0\}_\R\setminus (D_k)_\R$ is $(\frac{\partial G}{\partial X_1}\klk \frac{\partial G}{\partial X_k})$--regular.}\spar

{\it For any $1\le j \le s$ the semialgebraic set $(C_j)_\R$ is non empty. Moreover, any irreducible component of $W$ contains a real point.}\medskip

If Condition D is satisfied by $G$ we can prove in the same way as in case of Lemma \ref{l:1},  \ref{l:2} and \ref{l:4} the following corresponding statements.

\begin{lemma}\label{l:1'}
Let notations be as in Condition D, which we suppose to be satisfied. Then
for any $1\le j \le s$, we have $\dim C_j=n-k$.
\end{lemma}
\begin{lemma}\label{l:2'} 
Suppose that Condition D is satisfied and
let $C$ be an irreducible component of $W$. Then $G$ takes a constant real value on $C$.
\end{lemma}
Let $1\le k \le n$. By Lemma \ref{l:2'} the polynomial $G$ takes on $D_k$ only finitely many values which are all real algebraic. We denote by $B_k:=G(D_k)$ the set of these values. 
\begin{lemma}\label{l:4'}
Suppose that Condition D is satisfied.
Then the set of isolated local minimal points of $G$ in $\R^n$ is exactly the set of points of $W_\R$ where the Hessian of $G$ is positive definite.
\end{lemma}

Let $G\in \R[\xon]$ be a generic polynomial of degree two. The linear  subspaces defined by $\frac{\partial G}{\partial X_1}\klk \frac{\partial G}{\partial X_k}$ intersect transversally. Hence  Condition D is satisfied.

\subsection{Algorithms} \label{ss:2.2}
Let notations and assumptions be as in the previous subsection. We associate with $G$ and $\Fop$ intrinsic invariants that control the complexity of the algorithms we are going to develop in order to solve the computational problems of minimizing locally and globally $G$ on the set of points in $\R^n$ defined by the equality constraints $F_1=0\klk F_p=0$.\spar

Let $G$ and $\Fop\in \Q[\xon]$ be given as outputs of an essentially
division--free arithmetic circuit $\beta$ in $\Q$ having size $L$.
Let $d\ge 2$ be an upper bound for $\deg G, \deg F_1\klk \deg F_p$.

\subsubsection{The isolated local minimal point searching problem}\label{sss:2.2.1}
In this subsection we shall assume that the polynomials $G$ and $\Fop$ satisfy Condition B and C.\spar

We consider the task of finding all isolated local minimal points of $G$ in $V_\R=\{F_1=0\klk F_p=0\}_\R$. For this purpose we search for every index sequence $\J$ the isolated local minimal points of $G$ in the corresponding chart $(V_\R)_{\Delta_\J}$. Let $R_\J$ be the determinant of the Jacobian of $F_1\klk F_p, M_1^\J\klk M_{n-p}^\J$ and let $\delta_\J$ be the maximal degree of the Zariski closures in $\C^n$ of all locally closed sets
\begin{multline*}
\{F_1=0\klk F_j=0\}_{\Delta_\J\cdot R_\J},\;\;1\le j \le p,\;\;\text{and}\\
\{F_1=0\klk F_p=0,\, M_1^\J=0\klk M_k^\J=0\}_{\Delta_\J\cdot R_\J},\; 1\le k \le n-p.
\end{multline*}

Let finally
\[\delta:=\max \{\delta_\J\;|\;\J \;\;\text{index sequence} \}.\]
We call $\delta$ the {\it degree} of the isolated minimum searching problem for $G$ on $V_\R=\{F_1=0\klk F_p=0\}_\R$. From the B\'ezout Inequality we deduce
\[\delta \le (n\,d)^{O(n)}.\]
Fix for the moment  an index sequence $\J$ and observe that the polynomials \linebreak
$F_1\klk F_p, M_1^\J\klk M_{n-p}^\J$ generate the trivial ideal or form a reduced regular sequence in $\Q[\xon]_{\Delta_\J\cdot R_\J}$. Therefore we may apply
the Kronecker algorithm \cite{gilesa}, Theorem 1 and 2,  to the input system
\[
F_1=0\klk F_p=0,\, M_1^\J=0\klk M_{n-p}^\J=0,\;\;\Delta_\J\cdot R_\J\neq 0\]
in order to obtain for the complex points of 
\[\{F_1=0\klk F_p=0,\, M_1^\J=0\klk M_{n-p}^\J=0\}_{\Delta_\J\cdot R_\J}\]
 an algebraic description by univariate polynomials over $\Q$. There are at most $\delta_\J$ such points. For the real points among them we obtain even a description \`a la Thom. We now discard the points with non-zero imaginary part and
evaluate the signature of the Hessian matrix $H_\J$ at each of the real points and discard the real points where the Hessian is not positive definite. The remaining real points are by Lemma \ref{l:4} exactly the isolated local minimal points of $G$ in $(V_\R)_{\Delta_\J}$. Repeating this procedure for each index sequence we obtain all isolated local minimal points of $G$ in $V_\R$.
\medskip

The complexity analysis of \cite{gilesa}, Theorem 1 and 2, yields a time bound of
$L\,\binom{n}{p} (n\,d)^{O(1)}\,\delta^2$ for the first, algebraic part of the procedure. 
The sign evaluations necessary to handle real algebraic points make increase the overall complexity to $L\,\binom{n}{p} (n\,d)^{O(1)}\,\delta^3\le (n\,d)^{O(n)}$.
Applying \cite{bghlms}, Lemma 10 in the spirit of \cite{bghlms}, Section 5.1,
one can show that one may find probabilistically regular matrices
$A_1\klk A_{n-p+1}\in \Z^{n\times n}$ of logarithmic heights $O(n
\log dn)$ and $p$--minors $\bfs{\Delta}_1\klk \bfs{\Delta}_{n-p+1}$
of $J\iFop\cdot A_1\klk J\iFop\cdot A_{n-p+1}$
 such that $V_{\bfs{\Delta}_1}\cup \cdots \cup V_{\bfs{\Delta}_{n-p+1}}$ is the regular locus of $V$. Thus we may improve the sequential bound above to $L\,(n\,d)^{O(1)}\,\delta^3$.

\begin{theorem}\label{t:2}
Let $n,\,d,\,p,\,L,\,\,\delta\in \N$ with $1\le p \le n$ be
arbitrary and let $G,F_1\klk F_p\in \Q[\xon]$ be polynomials of
degree at most $d$ satisfying Condition B and C and having isolated
local minimum searching degree at most $\delta$. Suppose that $G,
F_1\klk F_p$ are given as outputs of an essentially division-free
circuit $\beta$ in $\Q[\xon]$ of size $L$.\medskip

There exists a uniform bounded error probabilistic algorithm
$\mathcal{B}$ over $\Q$ which computes from the input $\beta$ in
time $L\,(nd)^{O(1)}\delta^3 \le(nd)^{O(n)}$ all isolated local minimal
points of $G$ in $V_\R=\{F_1=0\klk F_p=0\}_\R$.
\medskip

For any $n,\,d,\,p,\,L,\,\delta \in \N$ with $1\le p \le n$
the probabilistic algorithm $\mathcal{B}$ may be realized by an
algebraic computation tree over $\Q$ of depth $L\,(n\,
d)^{O(1)}\,\delta^3\le (n\,d)^{O(n)}$ that depends on certain parameters which
are chosen randomly.
\end{theorem}
\paragraph{The unrestricted case.}\label{sss:2.2.1.1}
Let us now consider the problem of searching for the isolated local minimal points in the case of unconstrained optimization. For this purpose we assume that the polynomial $G$ satisfies the Condition D.
Let $H$ be the Hessian matrix of $G$. Observe that the polynomials $\frac{\partial G}{\partial X_1}\klk \frac{\partial G}{\partial X_n}$ generate the trivial ideal or form a reduced regular sequence in $\Q[\xon]_{\det H}$. Let $\delta$ be the maximal degree of the Zariski closure in $\C^n$ of the locally closed sets $\{\frac{\partial G}{\partial X_1}=0\klk \frac{\partial G}{\partial X_k}=0\}_{\det H}$, $1\le k \le n$. We call $\delta$ the {\it degree} of the isolated minimum searching problem for $G$ on $\R^n$. The B\'ezout inequality implies $\delta\le (d-1)^n\le d^n$. Applying the Kronecker algorithm to the input system
\[
\frac{\partial G}{\partial X_1}=0\klk \frac{\partial G}{\partial X_n}=0,\;\; \det H\neq 0
\]
we obtain an analogous statement to Theorem \ref{t:2} for the isolated minimum searching problem in the unconstrained case with Condition B and C  replaced by Condition D.

\subsubsection{The global minimal point searching problem}\label{sss:2.2.2}

In this subsection we shall assume that the polynomials  $G$ and $\Fop$ satisfy Condition B.
The aim of the next algorithm is to compute a real algebraic point which is a global minimizer of $G$ in $V_\R=\{F_1=0\klk F_p=0\}_\R$ if there exists one.\spar
Let $x$ be a minimal point of $G$ in $V_\R$ and let $b:=G(x)$. Then $x$ belongs to $W$ and therefore there exists by Lemma \ref{l:2} an irreducible component of $W$ where $G$ takes only the value $b$. This fact will guarantee that we are able to find a minimizer of $G$ on $V_\R$. For  an index sequence $\J$ and an index $1\le k \le n-p$ let $\delta_{\J,k}$ be the maximal degree of the Zariski closures in $\C^n$ of all locally closed sets
\begin{multline*}
\{F_1=0\klk F_j=0\}_{\Delta_\J},\;\;1\le j \le p,\;\;\text{and}\\
\{F_1=0\klk F_p=0,\, M_1^\J=0\klk M_{k'}^\J=0\}_{\Delta_\J},\; 1\le k' \le k
\end{multline*}
and all generic dual polar varieties of
\[
\{F_1=0\klk F_p=0,\, M_1^\J=0\klk M_k^\J=0\}_{\Delta_\J}.
\]
Let finally
\[
\delta:=\max\{\delta_{\J,k}\;|\;\J\;\;\text{index sequence},\;\; 1\le k \le n-p\}.
\]
We call $\delta$ the {\it degree} of the global minimum searching problem for $G$ on $V_\R$. From the B\'ezout inequality we deduce
\[
\delta \le (n\,d)^{O(n)}.
\]
We construct now recursively in $1\le k \le n-p$ an ascending chain of finite sets $Y_k$ of real algebraic points of $V_\R$
such that $G(Y_k)$ contains the set $B_{k+1}:=G(D_{k+1})$ (see Subsection \ref{ss:2.1} for the definition of $B_{k+1}$).\medskip

In order to construct $Y_1$ we apply for any index sequence $\J$ the algorithm of Theorem \ref{t:0} to the system $F_1=0\klk F_p=0,\, \Delta_\J\neq 0$. The algorithm returns a finite set $Y_1$ of algebraic points of $V_\R$.\medskip

Let now $2\le k \le n-p$ and suppose that we have already constructed $Y_{k-1}$
subject to the condition $B_k\subset G(Y_{k-1})$.\spar
We apply now for any index sequence $\J$ the same algorithm to the system
\[
F_1=0\klk F_p=0,\, M_1^\J=0\klk M_k^\J=0, \Delta_\J \neq 0.
\]
In this way we obtain finitely many algebraic points of $V_\R$ which together with
$Y_{k-1}$ form $Y_k$.\spar
Let us consider an arbitrary irreducible component $C$ of $W$ of dimension $n-p-k$ on which, by virtue of Lemma \ref{l:2}, the constant value of $G$ does not belong to $G(Y_{k-1})$. Then $B_k\subset G(Y_{k-1})$ implies $C\cap D_k=\emptyset$.\spar
 
Let us fix a generic vector $a :=(a_0, a_1\klk a_n)$ of $\Q^{n+1}$. Since 
$C_\R$ is closed, there exists a point $x$ of $C_\R$ which realizes the distance of
$(\frac{a_1}{a_0}\klk \frac{a_n}{a_0})$ to $C_\R$. The point $x$ belongs to $V_\R$
and therefore there exists an index sequence $\J$  with $\Delta_\J(x)\neq 0$. From $C\cap D_k=\emptyset$ we deduce $x\notin D_k$ and from $\Delta_\J(x)\neq 0$ and $x\in W$ we conclude
$x\in V_{\Delta_\J}\cap \{M_1^\J=0\klk M_k^\J=0\}$. Hence $x$ belongs to
\[(V_\R)_{\Delta_\J}\cap \{M_1^\J=0\klk M_k^\J=0\}_\R\setminus (D_k)_\R\]
and Condition B implies now that $x$ is $(F_1\klk F_p, M_1^\J\klk M_k^\J)$--regular. Therefore there exists just one irreducible component of 
$\{F_1=0\klk F_p=0, M_1^\J=0\klk M_k^\J=0\}$ which passes through $x$. This component is necessarily of dimension $n-p-k$ and contains $C$. It is therefore identical with $C$. Putting all this information together, we conclude that $x$ is a local minimizer of the distances of  $(\frac{a_1}{a_0}\klk \frac{a_n}{a_0})$ to the points of the real trace of $\{F_1=0\klk F_p=0, M_1^\J=0\klk M_k^\J=0\}$.
The point $x$ belongs therefore to the $(n-p-k)$th generic dual polar variety of
$\{F_1=0\klk F_p=0, M_1^\J=0\klk M_k^\J=0\}$ associated with $a$. Hence $x$ becomes computed by our algorithm. This implies $x\in Y_k$. Since $C$ was an arbitrary irreducible component of $W$ of dimension $n-p-k$
on which the constant value of $G$ does not belong to $G(Y_{k-1})$, we conclude
that $G(Y_k)$ contains the set $B_{k+1}$.\spar

Applying this argument inductively we see that $G(W)\subset G(Y_{n-p})$ holds. We suppose now that $G$ reaches a global minimum on $\{F_1=0\klk F_p=0\}_\R$.
Then $Y_{n-p}$ must contain a global minimal point of $G$ in $V_\R=\{F_1=0\klk F_p=0\}_\R$ which is an element, say $y$, of  $Y_{n-p}$ with $G(y)=\min_{x\in Y_{n-p}}G(x)$.
\medskip

The complexity analysis of the algorithm of  Theorem \ref{t:0} yields a time bound of\linebreak
$L\binom{n}{p}(n\,d)^{O(1)}\delta^2$ for the first algebraic part of the
procedure. The sign evaluations necessary to handle real algebraic points make increase the overall complexity to $L\,\binom{n}{p} (n\,d)^{O(1)}\,\delta^3\le (n\,d)^{O(n)}$.
We may use the same argumentation as in Theorem
\ref{t:2} in order to improve this bound to
$L\,(n\,d)^{O(1)}\,\delta^3$. We obtain now the following complexity
result.

\begin{theorem}\label{t:3}
Let $n,d,p,L,\,\delta\in \N$ with $1\le p \le n$ be
arbitrary and let $G,F_1\klk F_p\in \Q[\xon]$ be polynomials of
degree at most $d$ satisfying Condition B and having
global minimum searching degree at most $\delta$. Suppose that $G,
F_1\klk F_p$ are given as outputs of an essentially division-free
circuit $\beta$ in $\Q[\xon]$ of size $L$.\medskip

There exists a uniform bounded error probabilistic algorithm
$\mathcal{C}$ over $\Q$ which computes from the input $\beta$ in
time $L\,(nd)^{O(1)}\,\delta^3\le (nd)^{O(n)}$ a global minimal point
of $G$ in $V_\R=\{F_1=0\klk F_p=0\}_\R$ if there exists one.
\medskip

For any $n,\,d,\,p,\,L,\,\,\delta \in \N$ with $1\le p \le n$
the probabilistic algorithm $\mathcal{C}$ may be realized by an
algebraic computation tree over $\Q$ of depth
$L\,(nd)^{O(1)}\delta^3\le (nd)^{O(n)}$ that depends on certain parameters
which are chosen randomly.
\end{theorem}

Mutatis mutandis, with Condition D replacing Condition B, the same statement holds true for the unconstrained optimization problem. The degree of the minimum searching problem in this case is the maximal degree 
the closed sets $\{\frac{\partial G}{\partial X_1}=0\klk \frac{\partial G}{\partial X_k}=0\}\;1\le k \le n$ and all generic dual polar varieties of them.
\medskip

The reader should observe that Theorem \ref{t:3} does not answer the question whether $G$ reaches a global minimum on $\{F_1=0\klk F_p=0\}_\R$ and can only be applied when this existence problem is already solved. For this problem we refer to \cite{gremo}.

\section{Conclusion}

Together with \cite{legum} this paper represents only a first attempt to introduce the viewpoint of intrinsic quasi--polynomial complexity to the field of polynomial optimization. For this purpose we used some restrictive conditions which allow us to apply our algorithmic tools. In the future we shall relax these restrictions and extend the algorithmic tools and the list of real problems which can be treated in this way.
\medskip

Here we want to point to another modern approach to global polynomial optimization based on the so called relaxation which reduces the task under consideration to semi-definite programming. As in our situation this method requires that certain conditions, which involve the Karush--Kuhn--Tucker ideal, become satisfied. Moreover, no complexity bounds are available at this moment for this method. On the other hand we rely on tools of real polynomial equation solving which restricts the generality of our complexity results. For details about the relaxation approach we refer to \cite{las},  \cite{denipo}, \cite{nie}, \cite{ha} and \cite{bumo}.


\end{document}

r